\documentclass[12pt,fleqn]{article}
\usepackage[leqno]{amsmath}
\usepackage{amssymb, amsfonts}
\usepackage{geometry}
\usepackage{setspace}
\usepackage{indentfirst}
\usepackage{hyperref}
\usepackage{graphicx}
\usepackage{caption}
\usepackage{natbib}  
\begin{document}

\begin{center}
{\LARGE Transforming Investment Strategies and Strategic Decision-Making: Unveiling a Novel Methodology for Enhanced Performance and Risk Management in Financial Markets\

} 
\vspace{15pt}
Tian Tian,  Ricky Cooper, Jiahao Deng, and Qingquan Zhang\textsuperscript{*} \\
\vspace{10pt}
\end{center}

\noindent \textbf{Abstract:}This paper introduces a novel methodology for index return forecasting, blending highly correlated stock prices, advanced deep learning techniques, and intricate factor integration. Departing from conventional cap-weighted approaches, our innovative framework promises to reimagine traditional methodologies, offering heightened diversification, amplified performance capture, and nuanced market depiction. At its core lies the intricate identification of highly correlated company clusters, fueling predictive accuracy and robustness. By harnessing these interconnected constellations, we unlock a profound comprehension of market dynamics, bestowing both investment entities and individual enterprises with invaluable performance insights. Moreover, our methodology integrates pivotal factors such as indexes and ETFs, seamlessly woven with Hierarchical Risk Parity (HRP) portfolio optimization, to elevate performance and fortify risk management. This comprehensive amalgamation refines risk diversification, fortifying portfolio resilience against turbulent market forces. The implications reverberate resoundingly. Investment entities stand poised to calibrate against competitors with surgical precision, tactically sidestepping industry-specific pitfalls, and sculpting bespoke investment strategies to capitalize on market fluctuations. Concurrently, individual enterprises find empowerment in aligning strategic endeavors with market trajectories, discerning key competitors, and navigating volatility with steadfast resilience. In essence, this research marks a pivotal moment in economic discourse, unveiling novel methodologies poised to redefine decision-making paradigms and elevate performance benchmarks for both investment entities and individual enterprises navigating the intricate tapestry of financial realms.

\vfill

\noindent\textsuperscript{*}Tian Tian, Ph.D.In Management Science, a Principal AI Application Scientist excels in advanced AI techniques and cutting-edge sensitivity research. Her specialization is quantitative analysis, focusing on AI-driven solutions for industry advancement.Ricky Cooper, Full Professor of Finance at Illinois Institute of Technology.  Jiahao Deng, Ph.D. in Computer Science. His primary area of expertise lies in data visualization and machine learning.  Prof. Qingquan Zhang is an Assistant Professor of Finance at the University of Illinois at Urbana-Champaign with a Grand of Disruption and Innovation Fellow at the University of Illinois, 2021-2022
\newpage


\section*{I. Introduction}
In the ever-evolving investment landscape, constructing index returns using highly correlated stock prices has gained considerable significance\citet{1,2,3}. This approach offers numerous benefits and has become a cornerstone for various investment strategies and analysis methodologies. Moreover, recent advancements in data science and machine learning techniques have further expanded the possibilities for predicting and optimizing index returns, leading to novel and promising approaches \citet{4,5}.

\subsection*{A. Novelties of the approach}
Firstly, building index returns based on highly correlated stock prices allows companies to benchmark themselves against competitors \citet{6,7}. By comparing their performance to industry-specific indexes, companies can gauge their relative standing and identify areas for improvement \citet{8}. This evaluation fosters healthy competition and provides valuable insights for strategic decision-making.

Secondly, investors seeking exposure to specific industries aim to eliminate risks associated with individual stocks \citet{9,10}. Constructing industry-focused index returns provides them with a diversified investment option that captures the sector's overall performance while reducing idiosyncratic risk \citet{11}. This facilitates portfolio diversification and aligns with the preferences of risk-averse investors \citet{12}.

Moreover, constructing index returns using highly correlated stocks enables better peer grouping for fundamental analysis \citet{13,14,15,16}. Investors and analysts can assess industry performance and trends by analyzing the collective behavior of stocks within the index \citet{17}. This approach allows for more accurate comparisons and insights into sector-specific factors affecting returns, leading to informed investment decisions.

However, traditional time series techniques assume continuously changing processes in continuous time series data \citet{18,19}. In reality, time series data can exhibit non-continuous values, such as gaps or irregular intervals between observations. This discrepancy poses challenges for accurate forecasting using conventional methods \citet{20,21,22}. To overcome this limitation, deep learning models, specifically the combination of CNN and LSTM networks (CNN-LSTM), have emerged as a powerful approach. This hybrid model captures spatial and temporal patterns, extracts relevant features, and handles variable-length sequences, making it well-suited for predicting correlated index returns with non-continuous values \citet{23,24,25}.

Furthermore, integrating factors such as indexes and ETFs into index return forecasting has been shown to improve accuracy \citet{26}. These factors provide additional information and enhance model specification, allowing for a more comprehensive analysis of market dynamics and risk factors. Moreover, incorporating Hierarchical Risk Parity (HRP) portfolio optimization helps build industry indices by optimizing risk diversification \citet{27,28}. HRP considers the hierarchical structure of asset correlations, robustly handles noisy correlations, and adapts to changing market conditions, resulting in more accurate and stable industry indices.

\subsection*{B. Advantages over traditional cap-weighted methods}

The revolutionized method of index return forecasting, which leverages highly correlated stock prices, factors, and cutting-edge deep learning techniques, offers several advantages over traditional cap-weighted methods. Here are some reasons why this method is superior:
Improved Diversification: Unlike cap-weighted methods that may result in concentrated holdings of a few large stocks, the revolutionized method emphasizes highly correlated stocks. This approach leads to better diversification across the index, reducing the impact of individual stock performance and potentially mitigating risk.

 \underline{Enhanced Performance Capture}: By incorporating factors such as indexes and ETFs, the revolutionized method captures additional information and market dynamics. It allows for a more comprehensive analysis of industry-specific factors that impact returns. In contrast, cap-weighted methods solely rely on market capitalization, potentially overlooking critical insights.\

 \underline{More Accurate Market Representation}: The revolutionized method aligns with fundamental analysis by considering highly correlated stocks and factors. This provides a more accurate representation of the underlying market dynamics and allows for a deeper understanding of the industry's performance. Cap-weighted methods may not adequately capture these nuances, leading to potential misinterpretation of the market's true state.

 \underline{Adaptability to Non-Continuous Data}: Traditional time series techniques assume continuous data, which may not accurately reflect real-world scenarios. The deep learning models, such as CNN-LSTM, used in the revolutionized method are designed to handle non-continuous values and irregular intervals between observations. This enables more accurate forecasting and prediction of correlated index returns.

 \underline{Robust Risk Management}: Incorporating Hierarchical Risk Parity (HRP) portfolio optimization ensures optimal risk diversification when constructing industry indices. HRP considers the hierarchical structure of asset correlations, handles noisy correlations effectively, and adapts to changing market conditions. This leads to more stable and robust industry indices compared to cap-weighted methods, which may be more susceptible to extreme price movements of individual stocks.

 \underline{Future-Proofing}: The revolutionized method embraces cutting-edge deep learning techniques and factors, making it more adaptable to emerging trends and advancements in the investment landscape. It allows for the incorporation of new data sources and methodologies, ensuring relevance and effectiveness in an ever-changing market environment.

\subsection*{C. Structure of the paper}
In conclusion, the construction and prediction of index returns have witnessed significant advancements in recent years. The revolutionized method of index return forecasting, with its emphasis on correlated stocks, factors, and deep learning techniques, outperforms cap-weighted methods by offering improved diversification, enhanced performance capture, accurate market representation, adaptability to non-continuous data, robust risk management, and future-proofing capabilities. These advantages empower investors, companies, and analysts to make better-informed decisions and navigate the complexities of the financial world with greater precision and confidence. Section 2 describes dataset building and leverages highly correlated stock prices, incorporating factors, Section 3 employs HRP portfolio optimization to generate an industry index, and Section 4 is the prediction part to utilize CNN-LSTM models that have revolutionized the accuracy and effectiveness of index return forecasting. Section 5 is the conclusion and future study. These novel approaches empower companies, investors, and analysts with powerful tools to assess performance, manage risks, make informed investment decisions, and navigate the complex world of finance with greater precision.

\section*{II. Dataset Building}
\begin{figure}
\centering
\caption{Description of the datasets.}
\includegraphics[width=0.7\linewidth]{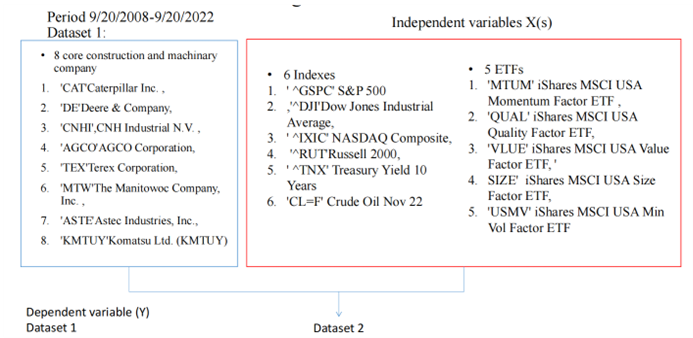}

\label{fig1}

\end{figure}
Figure \ref{fig1} presents the two datasets used in this research to investigate the prediction performance of the expected return of the industry index. Dataset 1 exclusively consists of historical data for the industry index return. On the other hand, Dataset 2 incorporates additional information by including six indexes and five ETFs alongside the industry index data. The purpose of incorporating these factors is to enhance the prediction performance of the expected return of the industry index.

Dataset 1 serves as the baseline, allowing for an evaluation of the predictive capabilities using only the historical data of the industry index return. By isolating the industry index data, the research aims to assess the inherent predictive power of this information alone.
In contrast, Dataset 2 introduces an expanded set of features by including six indexes and five ETFs alongside the industry index data. These additional factors capture broader market trends, sector-specific performance, and macroeconomic indicators. By incorporating this supplemental information, the research aims to investigate whether the prediction performance of the expected return of the industry index can be improved.

The comparison between Dataset 1 and Dataset 2 enables a comprehensive analysis of the impact of incorporating indexes and ETFs on the prediction accuracy of the industry index return. By evaluating the performance of different prediction models on both datasets, valuable insights can be gained regarding the significance and contribution of these additional factors in enhancing the prediction performance of the expected return of the industry index.
\subsection*{A. Dataset 1}
\subsubsection*{1. Reasons of choosing heavy machinery industry as an example to build industry index}
Choosing the heavy machinery industry as an example for building an industry index provides valuable insights into the broader economy, investment trends, and industry-specific dynamics \cite{29,32}. It allows for a comprehensive analysis of performance, risk assessment, and better decision-making within the industrial sector. Here are a few reasons why this industry is suitable for constructing an industry index:

\textbf{Economic Significance}: The heavy machinery industry plays a vital role in various sectors, such as construction, agriculture, mining, and infrastructure development \cite{31}. Tracking the performance of this industry provides valuable insights into the overall economic health and activity levels in these key sectors \cite{32}.

\textbf{Global Reach}: Heavy machinery companies often operate on a global scale, serving customers in multiple countries \cite{33}. Monitoring the performance of these companies helps gauge international market trends and provides a broader perspective on the global economic landscape \cite{34}.

\textbf{Investment and Capital Expenditure}: The heavy machinery industry involves significant capital investments and expenditures \cite{35}. Changes in industry performance and investment patterns can indicate shifts in business cycles, corporate spending, and investor sentiment, making it an essential sector to track for investment purposes \cite{36}.

\textbf{Cyclical Nature}: The heavy machinery industry is known for its cyclical patterns. Tracking the performance of companies in this industry can provide insights into economic cycles, such as periods of expansion, contraction, or recovery \cite{37,38}. This information can be valuable for investors, analysts, and policymakers in making informed decisions.

\textbf{Industry-specific Dynamics}: The heavy machinery industry has unique dynamics and factors that influence its performance. These may include commodity prices, government infrastructure spending, technological advancements, and environmental regulations \cite{39,40}. Building an industry index allows for a comprehensive analysis of these industry-specific factors and their impact on overall performance.

\textbf{Market Sentiment and Risk Assessment}: The heavy machinery industry is sensitive to changes in market sentiment, investor confidence, and global trade dynamics \cite{41,42}. By constructing an industry index, market participants can assess the risk profile, volatility, and potential returns associated with investing in heavy machinery companies.

Furthermore, the selection of specific companies to represent the heavy machinery industry index, based on those six criteria, was influenced by a combination of the following quantifiable metrics:
\begin{enumerate}
    \item \textbf{Economic Impact}: Evaluated using market capitalization, signaling each company's economic significance.
    \[ \text{Economic\_Impact}_i = \text{MarketCap}_i \]

    \item \textbf{Global Reach}: Based on international sales as a percentage of total sales.
    \[ \text{Global\_Reach}_i = \frac{\text{International\_Sales}_i}{\text{Total\_Sales}_i} \]

    \item \textbf{Capital Expenditure}: A reflection of their investment in the industry.
    \[ \text{Capital\_Expenditure}_i = \text{Annual\_CapEx}_i \]

    \item \textbf{Cyclical Nature}: Indicated by the stock's beta relative to the industry average.
    \[ \text{Beta}_i = \frac{\text{Cov}(R_i, R_m)}{\text{Var}(R_m)} \]

    \item \textbf{Industry-specific Dynamics}: Represented by key performance indicators (KPIs) pertinent to the heavy machinery industry.

    \item \textbf{Market Sentiment and Risk Assessment}: Measured by historical stock price volatility which is 
    \[ \text{Volatility}_i = \sigma(R_i) \]
\end{enumerate}
Note that 
\begin{equation}
R_i = \frac{\text{Ending Price of asset } i - \text{Beginning Price of asset } i + \text{Dividends from asset } i}{\text{Beginning Price of asset } i}
\end{equation} where
\begin{itemize}
    \item \( R_i \): Represents the return on asset \( i \) over a specified period. Daily value in our case.
    \item \textbf{Ending Price of asset } i: The price of the asset at the end of the period.
    \item \textbf{Beginning Price of asset } i: The price of the asset at the beginning of the period.
    \item \textbf{Dividends from asset } i: Any dividends or income received from the asset during the period.
\end{itemize}
The return, \( R_i \), represents the total gain or loss from an investment and is pivotal in our analysis. It captures both the price appreciation and dividends, providing a comprehensive view of a company's performance. This inclusion is particularly crucial in the heavy machinery industry, where dividends can form a significant component of returns. Additionally, the variability in returns aids in gauging the investment risk associated with each company, making it fundamental for understanding and weighing companies proportionally in the industry index. Furthermore, by considering both capital gains and dividends, the return serves as a barometer for the sector's overall economic condition, offering insights into the health of the industry and its implications for broader economic trends. Thus, integrating \( R_i \) into our dataset ensures a nuanced understanding of each company's contribution and the industry's economic standing.

Incorporating \( R_i \) into our analysis ensures a holistic and comprehensive evaluation of the heavy machinery industry's financial landscape.

\noindent Using a composite score derived from these metrics:

\begin{align*}
\text{Selection\_Score}_i = & w_1 \times \text{Economic\_Impact}_i \\
& + w_2 \times \text{Global\_Reach}_i \\
& + w_3 \times \text{Capital\_Expenditure}_i \\
& + w_4 \times |\text{Beta}_i - 1| \\
& + w_5 \times \text{KPI}_i \\
& + w_6 \times \text{Volatility}_i 
\end{align*}

\noindent the companies that best represented the industry were:

\begin{itemize}
    \item Caterpillar Inc. (CAT)
    \item Deere \& Company (DE)
    \item CNH Industrial N.V. (CNHI)
    \item AGCO Corporation (AGCO)
    \item Terex Corporation (TEX)
    \item The Manitowoc Company, Inc. (MTW)
    \item Astec Industries, Inc. (ASTE)
    \item Komatsu Ltd. (KMTUY)
\end{itemize}

\noindent We analyze daily data spanning the period 9/20/2008 to 9/20/2022. The weights $w_1$ to $w_6$ are user-defined parameters, with the sum of all weights equating to 1. The allocation of these weights determines the importance of each parameter. For our implementation, we have assigned equal values to all weights.

By combining these measurements of these companies, we can create an industry index that reflects the overall performance and trends within the machinery sector. This index will provide investors, analysts, and industry stakeholders with a benchmark to evaluate the performance of the industry as a whole.

The index will capture the aggregated movement of these companies' stock prices, representing the broader market sentiment and economic conditions impacting the heavy machinery industry. Constructing such an index allows for better industry analysis, risk assessment, and investment decision-making within the heavy machinery sector.

\subsubsection*{2. Highly correlated data characteristic of industry index portfolio built by the 8 core companies}

\begin{figure}
\centering
\caption{Correlation among 8 companies.}
\includegraphics[width=0.7\linewidth]{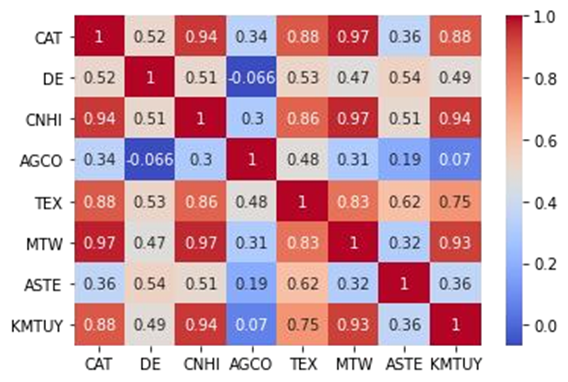}
\label{fig2}
\end{figure}

Constructing an industry index for the heavy machinery sector using eight core companies with highly correlated daily data from 9/20/2008 to 9/20/2022 ( showed by Figure \ref{fig2} correlation results) provides valuable insights into the overall performance and trends of this industry. By analyzing the characteristics of these companies' daily data, it is better to understand the dynamics and inter dependencies within the sector.

The high correlation among the daily data of these eight companies (showed by Fgiure \ref{fig2}) suggests that their stock prices move in tandem, indicating a strong association and synchronized performance within the heavy machinery industry. This correlation can be attributed to various factors, including shared market conditions, industry-specific events, macroeconomic influences, and investor sentiment.

Building a portfolio based on these highly correlated companies allows investors to gain exposure to the heavy machinery industry as a whole. It provides a diversified investment approach, spreading the risk across multiple companies, and potentially reducing the impact of company-specific events or volatility. By tracking the performance of this portfolio, investors can gauge the overall health and trends of the heavy machinery sector, enabling more informed investment decisions.

Moreover, the highly correlated nature of the daily data enhances the effectiveness of peer grouping and fundamental analysis within the industry. Comparing the performance, financial metrics, and key indicators among these companies becomes more meaningful and reliable when they exhibit strong correlation patterns. This correlation facilitates better benchmarking and evaluation of individual companies against their industry peers.

Finally, constructing an industry index for the heavy machinery sector using highly correlated daily data from these eight core companies provides a comprehensive and reliable measure of the sector's performance. It enables investors, analysts, and industry stakeholders to track and assess the industry's overall trends, identify investment opportunities, and make informed decisions based on the collective performance of these key players.

\subsection*{B. Factors added to build Dataset 2}

The selection of factors to help predict industry index performance is a critical step in developing an accurate forecasting model. The chosen factors should be relevant, representative, and have a significant impact on the industry in question. 

\subsubsection*{1. Six indexes added for predicting industry index performance}
Period: 9/20/2008 to 9/20/2022 daily data

\textbf{S\&P 500 (GSPC)}: The S\&P 500 is a widely recognized benchmark index that represents the performance of the largest publicly traded companies in the United States. Including the S\&P 500 allows for capturing broader market trends and general economic conditions that can influence the industry index.

\textbf{Dow Jones Industrial Average (DJI)}: The Dow Jones Industrial Average is another prominent index comprising 30 large, blue-chip companies. It provides insights into the performance of established and influential companies across various industries, offering a perspective on overall market sentiment and economic trends.

\textbf{NASDAQ Composite (IXIC)}: The NASDAQ Composite includes a wide range of technology and growth-oriented companies. Including this index accounts for the impact of technology-driven trends and innovation, particularly relevant in industries with a strong technological component.

\textbf{Russell 2000 (RUT)}: The Russell 2000 is a benchmark index that measures the performance of small-cap companies. Incorporating this index helps capture trends specific to smaller companies within the industry, as they often exhibit different growth patterns and market dynamics.

\textbf{Treasury Yield 10 Years (TNX)}: The yield on the 10-year Treasury note is an essential indicator of interest rates and serves as a barometer for market expectations. Changes in long-term interest rates can have significant implications for industries, particularly those sensitive to borrowing costs, such as capital-intensive sectors.

\textbf{Crude Oil (CL=F)}: Including crude oil as a factor accounts for the influence of energy prices on industries heavily dependent on oil and gas, such as transportation, manufacturing, and energy sectors. Fluctuations in crude oil prices can have a substantial impact on the profitability and performance of related industries.

By including these factors, the forecasting model can capture a wide range of economic, market, and industry-specific dynamics that affect the performance of the industry index. It provides a comprehensive perspective on both macroeconomic trends and factors specific to the industry being analyzed. This enhances the model's ability to predict and understand the performance of the industry index in response to various external influences.

\subsubsection*{2. Five ETFs added for predicting industry index performance}

Period: 9/20/2008 to 9/20/2022 daily data

\textbf{iShares MSCI USA Momentum Factor ETF (MTUM)}: This ETF focuses on stocks that exhibit strong price momentum. Including MTUM allows for capturing the performance of stocks experiencing upward price trends. Momentum is a crucial factor that can impact industry index performance, especially in industries driven by rapid growth and market sentiment.

\textbf{iShares MSCI USA Quality Factor ETF (QUAL)}: QUAL targets companies with strong financials, stable earnings, and high profitability. Incorporating QUAL helps assess the quality and financial stability of companies within the industry. The performance of companies with solid fundamentals often aligns with the overall performance of the industry index.

\textbf{iShares MSCI USA Value Factor ETF (VLUE)}: VLUE emphasizes stocks that are considered undervalued based on fundamental metrics such as price-to-earnings ratio and price-to-book ratio. Including VLUE provides insights into the performance of value-oriented companies within the industry. Value factors are particularly relevant for industries where investors seek out companies with attractive valuation metrics.

\textbf{iShares MSCI USA Size Factor ETF (SIZE)}: SIZE focuses on stocks of large-cap companies. Incorporating SIZE allows for capturing the performance of larger companies within the industry. Large-cap stocks often have a significant impact on industry indexes due to their market capitalization and influence.

\textbf{iShares MSCI USA Min Vol Factor ETF (USMV)}: USMV selects stocks with lower volatility and downside risk. Adding USMV helps account for the performance of stocks with more stable price movements within the industry. Low-volatility factors can be important for industries where investors prioritize stability and risk management.

By including these ETFs, the forecasting model can capture additional dimensions of market behavior and investor preferences that impact industry index performance. These ETFs represent different investment styles and factors that investors consider when constructing their portfolios. Incorporating them helps in understanding the underlying market dynamics and sentiment, providing a more comprehensive view of the industry index's predicted performance.

\section*{III. HRP(Hierarchical Risk Parity) portfolio optimization used to build industry index}
\begin{figure}
\centering
\caption{Markowitz mean-variance optimization result}
\includegraphics[width=0.7\linewidth]{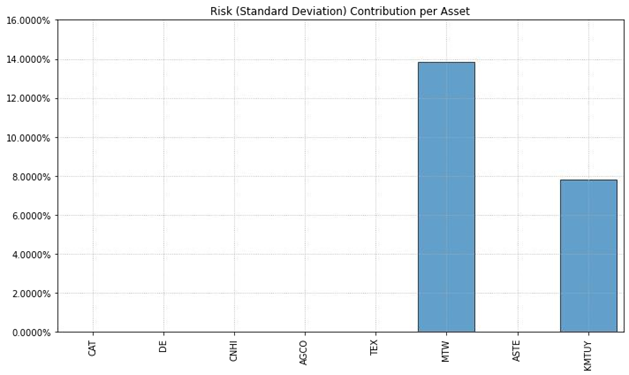}
\label{fig3}
\end{figure}
When it comes to allocating an industry index, the Hierarchical Risk Parity (HRP) approach stands out as a more suitable choice compared to Markowitz mean-variance optimization and other risk parity strategies such as full risk parity and naive risk parity.

\textbf{Markowitz's mean-variance optimization} often fails to capture the essence of building an industry index that tracks the entire industry's performance. In Figure \ref{fig3}, it is shown that Markowitz's mean-variance optimization may not necessarily select all the core companies in the index, which goes against the purpose of representing the industry as a whole.

To provide a mathematical description of this scenario, let's assume there are \( n \) core companies in the industry. Each company \( i \) (where \( i \) ranges from 1 to \( n \)) has an associated expected return, denoted by $\mu_i$, and a corresponding variance or volatility, denoted by $\sigma_i^2$.

In Markowitz's mean-variance optimization, the investor aims to construct a portfolio that maximizes returns for a given level of risk. This is typically achieved by finding the optimal weights, denoted by \( W_i \), to allocate to each asset in the portfolio. The weights satisfy the following constraints: 
\begin{align}
\sum W_i &= 1 && \text{(The weights must sum up to 1)} \\
W_i &\geq 0 && \text{(The weights must be non-negative)}
\end{align}
And the portfolio's variance, denoted by $\sigma_P^2$, is given by:
\begin{align}
    \sigma_P^2 = \sum \sum W_i W_j \sigma_i \sigma_j \rho_y && (\rho_y \text{ represents the correlation between assests i and j})
\end{align}
As Figure \ref{fig3}. shows, Markowitz's mean-variance optimization may not capture the essence of building an industry index that represents the entire industry when only a subset of core companies is chosen. Other factors, such as representativeness, and diversification, need to be considered to construct a comprehensive industry index.

\begin{figure}
\centering
\caption{Risk Parity weight allocated result}
\includegraphics[width=0.7\linewidth]{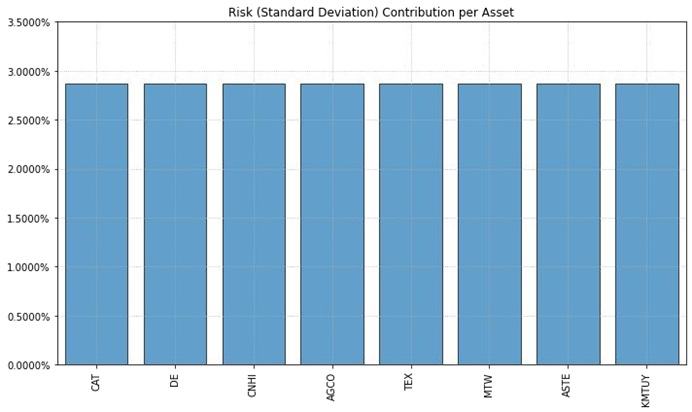}
\label{fig4}
\end{figure}
\textbf{Full risk parity}, as depicted in Figure \ref{fig4}, assigns equal weights to all companies regardless of their different characteristics. This approach overlooks the variations in the companies' importance and performance within the industry, potentially leading to an inaccurate representation of the industry index.To provide a mathematical and algorithmic expression for full risk parity, Let's assume there are n companies in the industry. Each company i (where i ranges from 1 to n) has an associated expected return, denoted by $\mu_i$, and a corresponding variance or volatility, denoted by $\sigma_i^2$.

In full risk parity, the algorithm aims to construct a portfolio where each company has an equal weight. The weights, denoted by $W_i$ , satisfy the following constraints:
\begin{align}
    W_i=1/n \text{(Equal weight assigned to each company)}
\end{align}
The portfolio's expected return, denoted by $\mu_P$ , is calculated as the weighted average of the expected returns of the individual companies:
\begin{align}
    \mu_P = \frac{\sum W_i \mu_i}{n}
\end{align}
The portfolio's variance, denoted by $\mu_P^2$ , can be calculated as:
\begin{align}
    \mu_P^2 = \frac{\sum W_i^2 \mu_i^2}{n}
\end{align}

\begin{figure}
\centering
\caption{HRP( Hierarchicial Risk Parity) optimization on weights allocated among 8 companies.}
\includegraphics[width=0.7\linewidth]{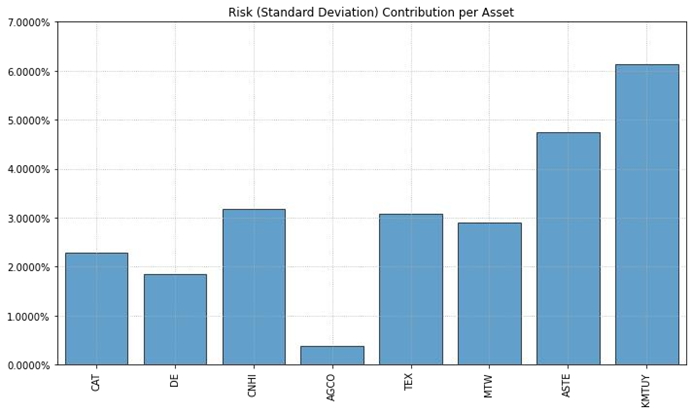}
\label{fig5}
\end{figure}

\textbf{HRP(Hierarchical Risk Parity)}, illustrated in Figure \ref{fig5}, takes into account the diverse characteristics of different companies and assigns them different weights while keeping all companies in the industry index for future tracking. This ensures a more comprehensive and representative index that accurately reflects the dynamics of the industry.
Let's define the variables used in the algorithm:
\begin{align*}
    n & : \text{The number of companies in the industry.} \\
    C & : \text{Covariance matrix (size } n \times n \text{).} \\
    D & : \text{Distance matrix (size } n \times n \text{).} \\
    L & : \text{Linkage matrix (size } n-1 \times 4 \text{).} \\
    W & : \text{Allocation weights vector (size } n \text{).}
\end{align*}
Compute the covariance matrix C based on historical returns of the companies.
Covariance matrix: A matrix denoting the pairwise covariance between the returns of the companies. Let's denote it as C (size n x n).
\\ 

\noindent Calculate the distance matrix:
Compute the distance matrix D (size n x n) based on the covariance matrix C using a chosen distance metric, such as the Euclidean distance.
\\

\noindent Construct the linkage matrix L using hierarchical clustering algorithms like single linkage, complete linkage, or Ward's method.\\

\begin{figure}
\centering
\caption{Weights allocated to each company}
\includegraphics[width=0.7\linewidth]{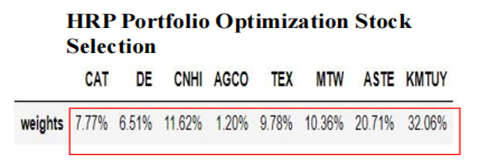}
\label{fig6}
\end{figure}

\noindent Calculate the allocation weights:
At each node i in the linkage matrix L:
\begin{itemize}
    \item Let k and j be the indices of the merged clusters at node i.
    \item Let $n_k$ and $n_j$ be the number of elements in clusters k and j, respectively.
    \item Calculate the inverse variance for the merged cluster at node \(i\) as
\[
\text{inverse\_variance}_i = \frac{1}{n_i \times n_j} \sum_{p \in \text{cluster } k}\sum_{q \in \text{cluster } j} C[p][q].
\]

    \item Allocate weights to each company in cluster \( k \) and cluster \( j \) proportionally to their inverse variances:
For each \( p \) in cluster \( k \), set \( W_p = \text{inverse\_variance}_i \times n_k \).
For each \( q \) in cluster \( j \), set \( W_q = \text{inverse\_variance}_i \times n_j \).

\end{itemize}
Repeat this step for each subsequent node in the dendrogram until weights are assigned to all companies. Normalize the weights vector \( W \) so that the sum of all weights equals 1.
Calculate the sum of weights: \( \text{sum}_w = \sum_{i=1}^{n} W_i \). 
Normalize each weight: \( W_i = \frac{W_i}{\text{sum}_w} \), for all \( i \) from 1 to \( n \).
By examining the weights shown in Figure \ref{fig6}, we can construct an industry index that appropriately captures the contributions of each company:
\begin{equation}
\begin{split}
\text{Expected}[\text{index\_return}] &= 0.0777 \times \text{['CAT']} + 0.0651 \times \text{['DE']} \\
&\phantom{=} + 0.1162 \times \text{['CNHI']} + 0.012 \times \text{['AGCO']} \\
&\phantom{=} + 0.0978 \times \text{['TEX']} + 0.2071 \times \text{['ASTE']} \\
&\phantom{=} + 0.1036 \times \text{['MTW']} + 0.3206 \times \text{['KMTUY']}
\end{split}
\end{equation}

\noindent This approach provides a more refined and effective allocation strategy, enabling the industry index better to track the overall performance of the heavy machinery industry. HRP offers a superior approach to allocating an industry index by considering the distinct characteristics of companies, maintaining a comprehensive

\subsection*{A. Reasons by choosing HRP}
 \underline{\textbf{Stability through Cluster Covariance}}: HRP replaces the individual covariances of assets with cluster covariances, which makes the portfolio more stable. By grouping assets into clusters based on their correlations, HRP considers the higher-level structure of the covariance matrix. This approach accounts for the inherent dependencies and similarities within the industry index, ensuring a more robust allocation.

Compute the cluster covariance matrix, denoted as $C_{\text{cluster}}$ (size $m \times m$), where $m$ is the number of clusters obtained from the hierarchical clustering.

The cluster covariance matrix $C_{\text{cluster}}$ is computed using the following equation:
\begin{align}
  C_{\text{cluster}}[i, j] = \sum_{p \in \text{cluster } i} \sum_{q \in \text{cluster } j} C[p, q]
\end{align}
where \( p \) and \( q \) are the indices of assets belonging to cluster \( i \) and cluster \( j \), respectively.

In other words, the cluster covariance between cluster \(i\) and cluster \(j\) is the sum of the individual covariances of assets within those clusters. Note that \(C_{\text{cluster}}\) is a square matrix of size \(m \times m\), where \(m\) is the number of clusters.

It's important to note that the cluster covariance matrix is derived from the hierarchical clustering and captures the aggregated covariance information between clusters. This aggregation helps to mitigate the impact of noise or idiosyncratic movements in individual asset covariances and promotes stability in the portfolio allocation process.

\underline{\textbf{Handling Noisy Covariances}}: 
Traditional mean-variance optimization, as used in Markowitz, relies heavily on accurate estimation of covariances. However, estimating precise covariances can be challenging due to noisy data and the instability of covariance estimates. HRP addresses this issue by reducing the impact of individual noisy covariances and focusing on the cluster-level relationships (matrix \(C_{\text{cluster}}\)). It offers a more reliable allocation strategy that is less susceptible to errors resulting from inaccurate covariance estimates.

\underline{\textbf{Improved Diversification}}: 
HRP's cluster-based approach enhances diversification within the industry index. By considering the hierarchical structure of asset correlations, HRP ensures that assets within each cluster are more closely related to each other, while assets across different clusters are less correlated. This leads to a more effective risk diversification and reduces the concentration risk in the portfolio, resulting in a better allocation strategy.

Compute the cluster-level average correlations, denoted as \(\text{avg\_corr\_cluster}\) (size \(m \times m\)), which represent the average correlation values between clusters.
The value \(\text{avg\_corr\_cluster}[i, j]\) represents the average correlation between assets in cluster \(i\) and assets in cluster \(j\). Compute the cross-cluster average correlations, denoted as \(\text{avg\_corr\_cross}\) (size \(m \times m\)), which represent the average correlation values between assets in different clusters.
\(\text{avg\_corr\_cross}[i, j]\) represents the average correlation between assets in cluster \(i\) and assets in cluster \(j\), where \(i \neq j\). Additionally, \(\text{avg\_corr\_cross}[i, i] = 0\) indicates no correlation between assets within the same cluster.

By considering the hierarchical structure of asset correlations, HRP ensures that assets within each cluster are more closely related to each other (higher \(\text{avg\_corr\_cluster}\) values), while assets across different clusters are less correlated (lower \(\text{avg\_corr\_cross}\) values). This leads to improved diversification within the industry index, as assets within the same cluster tend to move in a more synchronized manner, while assets in different clusters exhibit lower correlations and provide diversification benefits.

The cluster-level average correlations, \(\text{avg\_corr\_cluster}\), and cross-cluster average correlations, \(\text{avg\_corr\_cross}\), are utilized in subsequent steps of the HRP algorithm, such as calculating inverse variances for allocation weights and determining the portfolio weights for each company. By enhancing diversification through the cluster-based approach, HRP mitigates concentration risk and improves the allocation strategy by ensuring a more effective spread of risk across the industry index.

\underline{\textbf{Resilience to Extreme Events}}: HRP's cluster-based approach also helps in mitigating the impact of extreme events. By considering the dependencies at different levels, HRP diversifies the portfolio in a way that is more resilient to shocks and systemic risks. This makes it a suitable approach for allocating industry indices that may be exposed to sector-specific or market-wide disruptions.

Compute the correlation matrix based on the cluster covariance matrix, denoted as \( \text{Corr\_cluster} \) (size \( m \times m \)):

\begin{align}
\text{Corr\_cluster}[i,j] &= \frac{\text{C\_cluster}[i,j]}{\sqrt{\text{C\_cluster}[i,i] \cdot \text{C\_cluster}[j,j]}} 
\end{align}

\noindent where \( \text{C\_cluster}[i, j] \) is the covariance between assets in cluster \( i \) and cluster \( j \). \textit{Note:} \( \text{Corr\_cluster} \) is a matrix of pairwise correlations between clusters. The cluster covariance matrix \( \text{C\_cluster} \) and the correlation matrix \( \text{Corr\_cluster} \) capture the relationships between clusters, allowing HRP to account for sector-specific or market-wide disruptions. The average correlation within each cluster (\( \text{avg\_corr\_cluster} \)) and between different clusters (\( \text{avg\_corr\_cross} \)) are crucial in identifying the diversification potential and resilience to extreme events.

By considering the dependencies and correlations at different levels, HRP allocates the portfolio in a manner that accounts for potential shocks and systemic risks. This approach diversifies the portfolio across clusters, helping to mitigate the impact of extreme events and enhance the resilience of the industry index allocation strategy.

\subsection*{B. Adavantages of choosing HRP}
Choosing \textbf{HRP (Hierarchical Risk Parity)} portfolio optimization to build an industry index offers several advantages and considerations that make it an appealing approach. Here are some reasons why HRP may be chosen for constructing an industry index:

\begin{itemize}
    \item \textbf{Risk Diversification:} HRP portfolio optimization focuses on achieving optimal risk diversification by considering the hierarchical structure of asset correlations. It takes into account both the pairwise correlations between assets and the cluster-level correlations among groups of assets. By incorporating the hierarchical structure, HRP aims to allocate weights to assets in a way that maximizes diversification and reduces overall portfolio risk. This is particularly important when constructing an industry index to ensure that the index captures the risk characteristics of the industry effectively.
    
    \item \textbf{Robustness to Noisy Correlations:} HRP is known for its robustness to noisy or unreliable correlations. In financial markets, correlations between assets can be volatile and subject to measurement errors. HRP's hierarchical approach helps mitigate the impact of noisy correlations by considering the larger picture of cluster-level relationships rather than relying solely on individual asset correlations. This robustness is valuable when constructing an industry index, as it ensures the stability and reliability of the index's composition and performance.
    
    \item \textbf{Adaptability to Changing Market Conditions:} HRP is designed to adapt to changing market conditions, which is particularly relevant in the dynamic nature of the stock market and industries. As correlations and relationships between assets evolve over time, HRP's hierarchical approach allows for efficient portfolio rebalancing and reallocation. This adaptability ensures that the industry index remains aligned with the changing dynamics and risk profiles of the industry, providing a more accurate representation of industry performance.
    
    \item \textbf{Scalability:} HRP portfolio optimization is scalable and can handle a large number of assets, making it suitable for constructing industry indices that encompass a wide range of companies within the industry. With many industry indices covering numerous stocks, the ability of HRP to handle large-scale optimization helps ensure that the index captures the diversification and risk characteristics of the entire industry effectively.
    
    \item \textbf{Interpretability:} HRP provides a hierarchical structure that can offer interpretability and insights into the relationships between different assets and groups within an industry. This can be valuable for investors and analysts seeking to understand the underlying dynamics and interdependencies within the industry. The interpretability aspect of HRP enhances transparency and facilitates informed decision-making when constructing an industry index.
\end{itemize}

\section*{IV. Index Return Prediction}
\subsection*{A. LSTM and CNN-LSTM prediction results on Dataset 1}
\begin{figure}
\centering
\caption{LSTM 100 Epochs prediction result on dataset1 }
\includegraphics[width=0.9\linewidth]{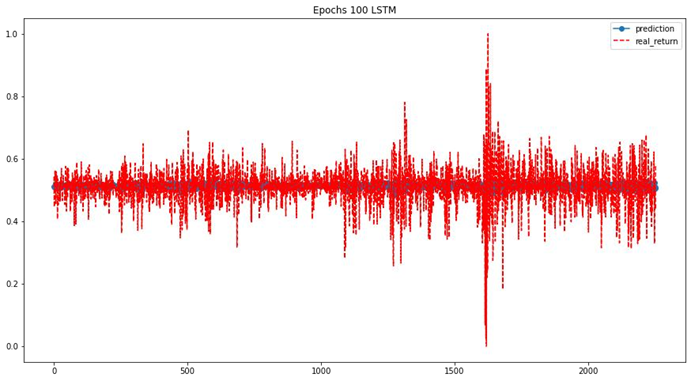}
\label{fig7}
\end{figure}

Despite the ability of Long Short-Term Memory (LSTM) networks to track previous data points and capture evolving patterns, using LSTM alone may not adequately capture the fluctuations in an industry index. In our study, we employed LSTM as the primary model for predicting index returns, as depicted in Figure \ref{fig7}. The results indicated that while LSTM could capture the overall trend of the index, it struggled to accurately capture the finer details of the fluctuations.

To evaluate the performance of the LSTM model, the research calculated the average Root Mean Square Error (RMSE) over 30 iterations, considering a range of $[0,1]$ for comparison purposes. The obtained average RMSE was $0.179$. This value reflects the discrepancy between the predicted and actual index returns and indicates the model's limitations in capturing the precise details of the fluctuations.

The findings suggest that relying solely on LSTM may not be sufficient for accurately predicting the subtle changes and fluctuations in an industry index. Additional factors or techniques should be considered to enhance the predictive capabilities. These could include incorporating external variables such as economic indicators, news sentiment analysis, or utilizing more advanced modeling approaches like ensemble methods to improve the model's performance and capture the finer nuances of the index fluctuations.

\textbf{CNN-LSTM prediction result}

\begin{figure}
\centering
\caption{LSTM index return prediction on dataset 1}
\includegraphics[width=0.9\linewidth]{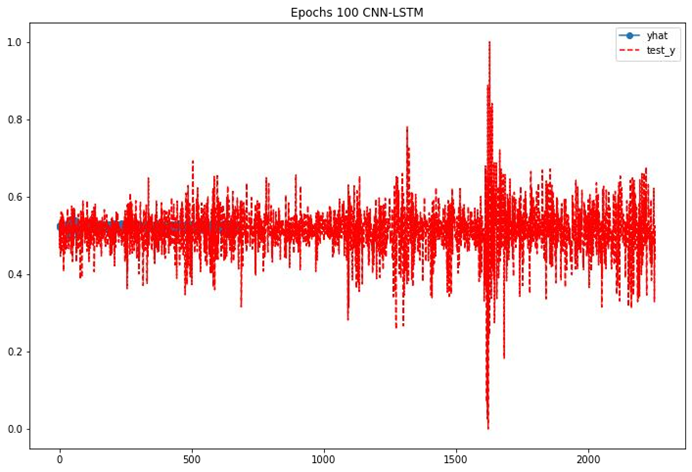}
\label{fig8}
\end{figure}

To improve the prediction results of LSTM, one approach is to incorporate Convolutional Neural Networks (CNN) as a preprocessing step by building a CNN-LSTM model. Through the average Root Mean Square Error (RMSE) over 30 iterations, considering a range of $[0,1]$ for comparison purposes. The obtained average RMSE was $0.088$ as shown in Figure \ref{fig8}. While this increases the prediction result of LSTM, the result still struggled to accurately capture the finer details of the fluctuations.

\subsection*{B. LSTM and CNN\_LSTM prediction results of Dataset 2 by adding factors }
\subsubsection*{1. Additional factors of 6 indexes and 5 ETFs added to improve the  prediction results significantly}
\begin{figure}
\centering
\caption{LSTM prediction result of Dataset 2}
\includegraphics[width=0.9\linewidth]{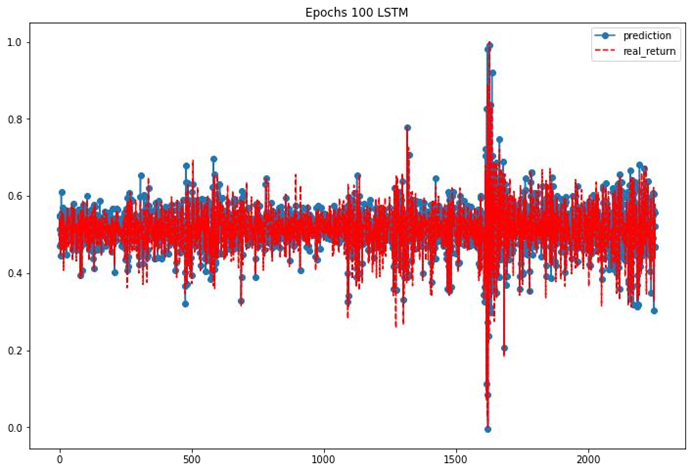}
\label{fig9}
\end{figure}

\begin{figure}
\centering
\caption{CNN-LSTM prediction result of Dataset 2}
\includegraphics[width=0.9\linewidth]{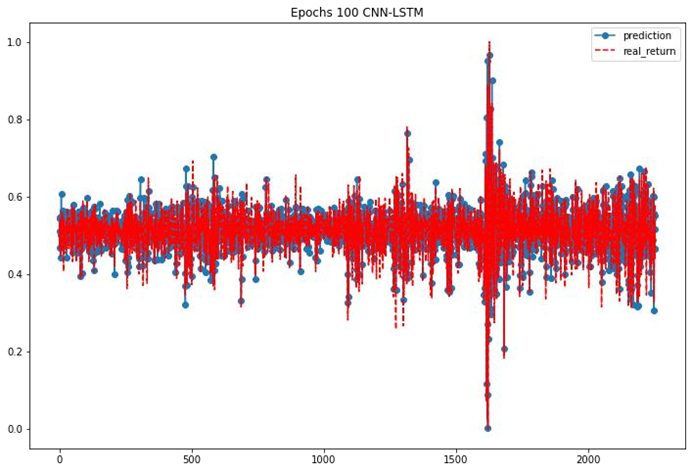}
\label{fig10}
\end{figure}

The prediction performance of LSTM and CNN-LSTM models was evaluated using Dataset 2, which includes the industry index data along with six indexes and five ETFs. Figure \ref{fig9} illustrates that the average root mean square error (RMSE) of LSTM predictions, based on 30 runs and 100 epochs, was $0.034$. This indicates a significant improvement compared to LSTM predictions using Dataset 1, which reduced the RMSE by $82.12\%$. Similarly, when utilizing the CNN-LSTM model with Dataset 2, the average RMSE, as shown in Figure \ref{fig10}, was $0.028$. This further reduced the RMSE by $84.35\%$ compared to LSTM predictions with Dataset 1.

The results demonstrate the effectiveness of incorporating additional factors, such as indexes and ETFs, in improving the prediction accuracy of the expected return of the industry index. The LSTM model performed considerably better when trained on Dataset 2, which suggests that the inclusion of these factors enhanced the model's ability to capture complex patterns and dependencies in the industry index data.

Furthermore, the CNN-LSTM model achieved even lower RMSE values, outperforming the LSTM model on both Dataset 1 and Dataset 2. With Dataset 2, the CNN-LSTM model achieved an average RMSE of $0.028$, reducing the RMSE by $68.19\%$ compared to the CNN-LSTM model trained on Dataset 1.

These findings emphasize the significance of incorporating additional factors, such as indexes and ETFs, in improving the prediction performance of industry index returns. The combination of LSTM and CNN-LSTM models with Dataset 2 demonstrates their effectiveness in capturing and leveraging the information provided by these factors, leading to more accurate predictions of the expected return of the industry index.

\subsubsection*{2. Reasons of incorporating additional factors can significantly enhance the prediction results}
Both LSTM and CNN-LSTM models may face limitations when making predictions solely based on daily data from the 8 core companies in the industry index. However, incorporating additional factors such as indexes and ETFs can significantly enhance the prediction results.

\begin{itemize}
    \item \textbf{Diversification of Information:} Including indexes and ETFs provides a broader perspective on the market and the industry as a whole. These factors capture the performance and trends of the overall market, specific sectors, or key economic indicators. By considering a wider range of data, the prediction model can better capture the macroeconomic and market influences on the industry index.

    \item \textbf{Enhanced Risk Management:} Indexes and ETFs represent a diversified portfolio of stocks within specific sectors or factors. By incorporating these factors, the prediction model can better manage specific risks associated with individual companies. This diversification helps reduce the impact of idiosyncratic factors and improves the robustness of the prediction model.

    \item \textbf{Improved Feature Representation:} The inclusion of indexes and ETFs introduces additional features that complement the information from the core companies' daily data. These features can capture market sentiment, industry performance, and macroeconomic factors that influence the industry index. The model can learn from these features and gain a deeper understanding of the complex relationships and dynamics impacting the index returns.

    \item \textbf{Expanding Data Availability:} Index and ETF data are widely available and typically represent reliable and accurate market information. By incorporating these factors, the prediction model benefits from a larger dataset, which can lead to more robust and accurate predictions.
\end{itemize}

By integrating indexes and ETFs into the prediction model, the prediction accuracy of the industry index returns improved significantly. These additional factors provide valuable insights into market trends, risk management, and overall market performance, enabling a more comprehensive analysis and prediction of the industry index.

\subsection*{C. Reasons why adding CNN before LSTM can be beneficial}
The CNN-LSTM model surpasses the LSTM model in terms of prediction performance on both Dataset 1 and Dataset 2. This superiority can be attributed to several factors related to the architecture and capabilities of CNNs and LSTMs.

CNNs are well-suited for extracting spatial features from input data, such as historical price or performance data of the industry index. By treating the data as a 2D image, CNNs can effectively capture spatial patterns and relationships. This spatial feature extraction is particularly valuable in understanding the complex dynamics within the industry index.

Moreover, CNNs excel at dimensionality reduction by applying filters and pooling operations. This enables them to extract high-level representations of the data while preserving essential features. Dimensionality reduction is crucial for handling large-scale datasets and identifying significant patterns within the industry index.

The hierarchical nature of feature extraction in CNNs complements the LSTM architecture. CNNs learn hierarchical representations of features, starting from low-level to high-level features. This hierarchy allows the subsequent LSTM layers to focus on capturing temporal dependencies and long-term patterns. By incorporating the CNN's extracted features as input sequences to LSTM, the model can better understand the underlying patterns in the industry index data.

Furthermore, the combination of CNN and LSTM leverages the strengths of both architectures, leading to enhanced representation learning. CNNs excel at spatial feature extraction, while LSTMs excel at capturing temporal dependencies. By integrating the two, the model can capture both spatial and temporal patterns simultaneously, resulting in improved prediction accuracy.

In summary, the addition of CNN as a preprocessing step before LSTM in the CNN-LSTM model enhances the model's ability to capture complex spatial patterns, reduce dimensionality, learn hierarchical features, and ultimately improve the prediction accuracy of the industry index.

\subsubsection*{1. Economic and financial implications of CNN-LSTM}
Indeed, the CNN-LSTM architecture, which combines Convolutional Neural Network (CNN) layers with Long Short-Term Memory (LSTM), can carry significant economic and financial implications, drawing upon concepts from game theory.

\begin{itemize}
    \item \textbf{Feature Extraction:} CNN layers in the architecture excel at extracting spatial features from input data, such as historical price or performance data. This feature extraction process helps identify relevant patterns and relationships within the data, similar to players in a game seeking to identify strategies or opportunities.
    
    \item \textbf{Sequence Prediction:} LSTM layers are effective in capturing temporal dependencies and predicting sequences. In the context of economic and financial data, LSTM can leverage its memory cells to understand long-term patterns and make predictions based on previous inputs, similar to players anticipating future moves based on past actions in a game.
    
    \item \textbf{Max(CNN) VS. Min(LSTM):} The combination of CNN and LSTM introduces a game-like dynamic to the architecture. CNN aims to maximize its ability to extract meaningful features from the data, while LSTM seeks to minimize prediction errors and capture long-term dependencies. This interplay between ``maximization'' and ``minimization'' aligns with the concept of players in a game seeking to optimize their strategies while minimizing potential losses.
    
    \item \textbf{Balance and Optimal Solutions:} The CNN-LSTM architecture seeks to strike a balance between feature extraction and sequence prediction, leveraging the strengths of both networks. This balance is similar to game theory, where players aim to find optimal solutions that maximize their outcomes within the constraints of the game.
\end{itemize}

By combining CNN and LSTM in the CNN-LSTM architecture, we create a framework that can effectively capture complex patterns, dependencies, and dynamics within economic and financial data. This architecture embodies the principles of game theory, with CNN and LSTM networks working together to achieve the best possible balance and predictive performance, ultimately enhancing our understanding and interpretation of economic and financial phenomena.

\section*{V. Conclusion and future study}
In conclusion, constructing index returns using highly correlated stock prices has gained significant importance in the investment landscape. It allows for benchmarking, risk mitigation, and better peer grouping for fundamental analysis. However, traditional time series techniques may not handle the non-continuous nature of real-world data. Deep learning models, particularly the CNN-LSTM architecture, have emerged as a powerful solution for predicting correlated index returns with non-continuous values.

Moreover, the incorporation of factors such as indexes and ETFs has proven to be crucial for improving the accuracy of index return forecasting. These additional factors capture valuable information and enhance the specification of the models, leading to more precise predictions.

Furthermore, Hierarchical Risk Parity (HRP) portfolio optimization provides an optimal approach for diversifying risks when constructing industry indices. By considering the hierarchical structure of asset correlations, HRP effectively handles noisy correlations and adapts to changing market conditions, resulting in a more stable and robust portfolio.

In summary, the combination of deep learning models like CNN-LSTM and the inclusion of factors like indexes and ETFs have revolutionized index return forecasting. These techniques have significantly enhanced the accuracy and effectiveness of predictions, allowing individuals to make informed investment decisions and navigate the complexities of the financial world with greater confidence. Emphasizing the importance of adding factors in the forecasting process highlights their value in improving the overall performance of the models.

Future studies in the field of index return forecasting can focus on several areas to further enhance the accuracy and effectiveness of predictions:

\begin{itemize}
    \item \textbf{Incorporating Additional Factors:} While this study considered the inclusion of indexes and ETFs, there may be other relevant factors that could impact industry index performance. Exploring and incorporating additional factors, such as macroeconomic indicators, sentiment analysis, or news sentiment, can provide a more comprehensive view and improve prediction accuracy.
    
    \item \textbf{Hybrid Models:} Investigating hybrid models that combine multiple techniques, such as ensemble models or hybrid architectures, can potentially yield better prediction results. Combining the strengths of different models, such as CNN, LSTM, or other deep learning architectures, along with traditional statistical models, can lead to more robust and accurate forecasts.
    
    \item \textbf{Feature Engineering:} Exploring advanced feature engineering techniques can uncover more meaningful patterns and relationships within the data. Feature selection, dimensionality reduction, and transformation methods can help identify the most relevant and informative features, enhancing the predictive power of the models.
    
    \item \textbf{Evaluation Metrics:} Developing and utilizing comprehensive evaluation metrics can provide a deeper understanding of model performance. Besides RMSE, other metrics such as Mean Absolute Percentage Error (MAPE), precision, recall, and F1-score can be employed to assess the models' ability to capture different aspects of index return fluctuations.
    
    \item \textbf{Incorporating External Data:} Incorporating external data sources, such as social media data, alternative data, or economic indicators, can enrich the predictive models. By integrating diverse data sources, models can capture real-time market dynamics and improve their forecasting capabilities.
    
    \item \textbf{Real-Time Forecasting:} Expanding the analysis to real-time forecasting can be valuable for market participants who require up-to-date predictions. Developing models that can adapt and update in real-time as new data becomes available can provide timely insights for investment decision-making.
    
    \item \textbf{Robustness and Stability Analysis:} Conducting robustness and stability analysis to assess the sensitivity of the models to different time periods, market conditions, or data variations can enhance their reliability and practical applicability.
\end{itemize}

By exploring these areas, future studies can further advance the field of index return forecasting and contribute to more accurate predictions, better risk management, and improved investment decision-making.



\newpage

\end{document}